\documentclass[prb,showpacs,preprintnumbers,amsmath,am'ssymb,twocolumn]{revtex4}

\usepackage{graphicx}
\usepackage{dcolumn}
\usepackage{bm}

\preprint{submitted to ``Spintronics: Materials, Applications and
Devices'', Nova Publishers}

\begin{document}

\title{Fundamentals of half-metallic Full-Heusler alloys}

\author{K. \"{O}zdo\u{g}an}\email{kozdogan@gyte.edu.tr}
\affiliation{Department of Physics, Gebze Institute of Technology,
Gebze, TR-41400, Kocaeli, Turkey}

\author{E. \c{S}a\c{s}{\i}o\u{g}lu}\email{e.sasioglu@fz-juelich.de}
\affiliation{Institut f\"{u}r Festk\"{o}rperforschung,
Forschungszentrum J\"{u}lich, D-52425 J\"{u}lich, Germany  \\ and
Fatih University,  Physics Department, TR-34500, B\"{u}y\"{u}k\c
cekmece, \.{I}stanbul, Turkey}

\author{I. Galanakis}\email{galanakis@upatras.gr}
\affiliation{Department of Materials Science, School of Natural
Sciences, University of Patras,  GR-26504 Patra, Greece}

\date{\today}

\begin{abstract}
Intermetallic Heusler alloys are amongst the most attractive
half-metallic systems due to the high Curie temperatures and the
structural similarity to the binary semiconductors. In this review
we present an overview of the basic electronic and magnetic
properties of the half-metallic full-Heusler alloys like
Co$_2$MnGe. Ab-initio results suggest that the electronic and
magnetic properties in these compounds are intrinsically related
to the appearance of the minority-spin gap.  The total spin
magnetic moment in the unit cell, $M_t$, scales linearly with the
number of the valence electrons, $Z_t$, such that $M_t=Z_t-24$ for
the full-Heusler alloys opening the way to engineer new
half-metallic alloys with the desired magnetic properties.
Moreover we present analytical results on the disorder in
Co$_2$Cr(Mn)Al(Si) alloys, which is susceptible to destroy the
perfect half-metallicity of the bulk compounds and thus degrade
the performance of devices. Finally we discuss the appearance of
the half-metallic ferrimagnetism due to the creation of Cr(Mn)
antisites in these compounds and the Co-doping in Mn$_2$VAl(Si)
alloys which leads to the fully-compensated half-metallic
ferrimagnetism.
\end{abstract}

\maketitle

\section{Introduction \label{sec1}}

Half-metallic ferromagnets, initially predicted by de Groot and
his collaborators in 1983\cite{deGroot} using first-principles
calculations are at the center of scientific research due to their
potential applications. The minority-spin electrons in these
compounds show a semiconducting electronic band-structure while
the majority-spin electrons present the usual metallic behavior of
ferromagnets. Thus such alloys would ideally exhibit a 100\% spin
polarization at the Fermi level and therefore they should have a
fully spin-polarized current and should be ideal spin injectors
into a semiconductor maximizing the efficiency of spintronic
devices.\cite{Zutic2004}

Heusler alloys are well-know for several decades and the first
Heusler compounds studied had the chemical formula  X$_2$YZ, where
X is a high valent transition or noble metal atom, Y a low-valent
transition metal atom and Z a sp element, and crystallized in the
$L2_1$ structure which consists of four fcc
sublattices.\cite{landolt,landolt2} Heusler compounds present a
series of diverse magnetic phenomena like itinerant and localized
magnetism, antiferromagnetism, helimagnetism, Pauli paramagnetism
or heavy-fermionic behavior and thus they offer the possibility to
study various phenomena in the same family of
alloys.\cite{landolt,landolt2} A second class of Heusler alloys
have the chemical formula XYZ and they crystallize in the $C1_b$
structure which consists of three fcc sublattices; they are often
called half- or semi-Heusler alloys in literature, while the
$L2_1$ compounds are referred to as full Heusler alloys. The
interest in these types of intermetallic alloys was revived after
the prediction,\cite{deGroot} using first-principles calculations,
of half-metallicity in NiMnSb, a semi-Heusler compound.

The high Curie temperatures\cite{landolt,landolt2} exhibited by
the half-metallic Heusler alloys are their main advantage with
respect to other half-metallic systems (e.g. some oxides like
CrO$_2$ and Fe$_3$O$_4$ and some manganites like
La$_{0.7}$Sr$_{0.3}$MnO$_3$).\cite{Soulen98} While for the other
compounds the Curie temperature is near the room temperature, for
the half-metal NiMnSb it is 730 K and for the half-metallic
Co$_2$MnSi it reaches the 985 K.\cite{landolt} A second advantage
of Heusler alloys for realistic applications is their structural
similarity to the zinc-blende structure, adopted by binary
semiconductors widely used in industry (such as GaAs on ZnS).
Semi-Heusler alloys have been already incorporated in several
devices as spin-filters,\cite{Kilian00} tunnel
junctions\cite{Tanaka99} and GMR devices.\cite{Caballero98}
Recently, also the half-metallic full-Heusler alloys have found
applications. The group of Westerholt has incorporated Co$_2$MnGe
in the case of spin-valves and multilayer
structures\cite{Westerholt} and the group of Reiss managed to
create magnetic tunnel junctions based on Co$_2$MnSi.
\cite{Reiss04,Reiss04b} A similar study by Sakuraba and
collaborators resulted in the fabrication of magnetic tunnel
junctions using Co$_2$MnSi as magnetic electrodes and Al-O as the
barrier and their results are consistent with the presence of
half-metallicity for Co$_2$MnSi.\cite{Sakuraba} Moreover Dong and
collaborators recently managed to inject spin-polarized current
from Co$_2$MnGe into a semiconducting structure.\cite{Dong}

 In the present contribution we review our most recent results on the electronic
properties of the half-metallic full-Heusler alloys obtained from
first-principles electronic structure calculations. In section
\ref{sec2} we summarize our older results on bulk compounds
obtained using the full-potential version of the screened
Korringa-Kohn-Rostoker (KKR) method\cite{KKR} (see references
\onlinecite{Review-JPD,Springer,Nova} for an extended review). In
the next sections we overview some of our results on the defects
in the half-metallic full Heusler alloys obtained using the
full--potential nonorthogonal local--orbital minimum--basis band
structure scheme (FPLO)\cite{koepernik1,koepernik2} within the
local density approximation (LDA) \cite{LDA1,LDA2,LDA3} and
employing the coherent potential approximation (CPA) to simulate
the disorder in a random way.\cite{koepernik2} In section
\ref{sec3} we present the physics of defects in the ferromagnetic
Heusler alloys containing Co and Mn like
Co$_2$MnSi.\cite{APL,PRB-def} The study of defects, doping and
disorder is of importance to accurately control the properties of
half-metallic full-Heusler alloys.\cite{PicozziReview} In section
\ref{sec4} we demonstrate the creation of half-metallic
ferrimagnets based on the creation of Cr and Mn antisites in
Co$_2$(Cr or Mn)(Al or Si) alloys\cite{PSS-RRL,SSC} and  we expand
this study to cover the case of Co defects in ferrimagnetic
Mn$_2$VAl and Mn$_2$VSi alloys leading to full-compensated
half-metallic ferrimagnets (also-known as half-metallic
antiferromagnets).\cite{PRB-HMA} Finally in section \ref{sec5} we
summarize and conclude.

\section{Electronic and gap properties -- Slater Pauling behavior}
\label{sec2}

The electronic, magnetic and gap properties of half-metallic
full-Heusler compounds have been studied using first-principles
calculations in reference \onlinecite{GalanakisFull}. These
results have been extensively also reviewed in reference
\onlinecite{Review-JPD}, in the introductory chapter of
\onlinecite{Springer} and in a chapter in reference
\onlinecite{Nova}. The reader is directed to them for an extended
discussion.  In this section we will only briefly overview these
properties.

\begin{figure}
\centering
\includegraphics[width=\linewidth]{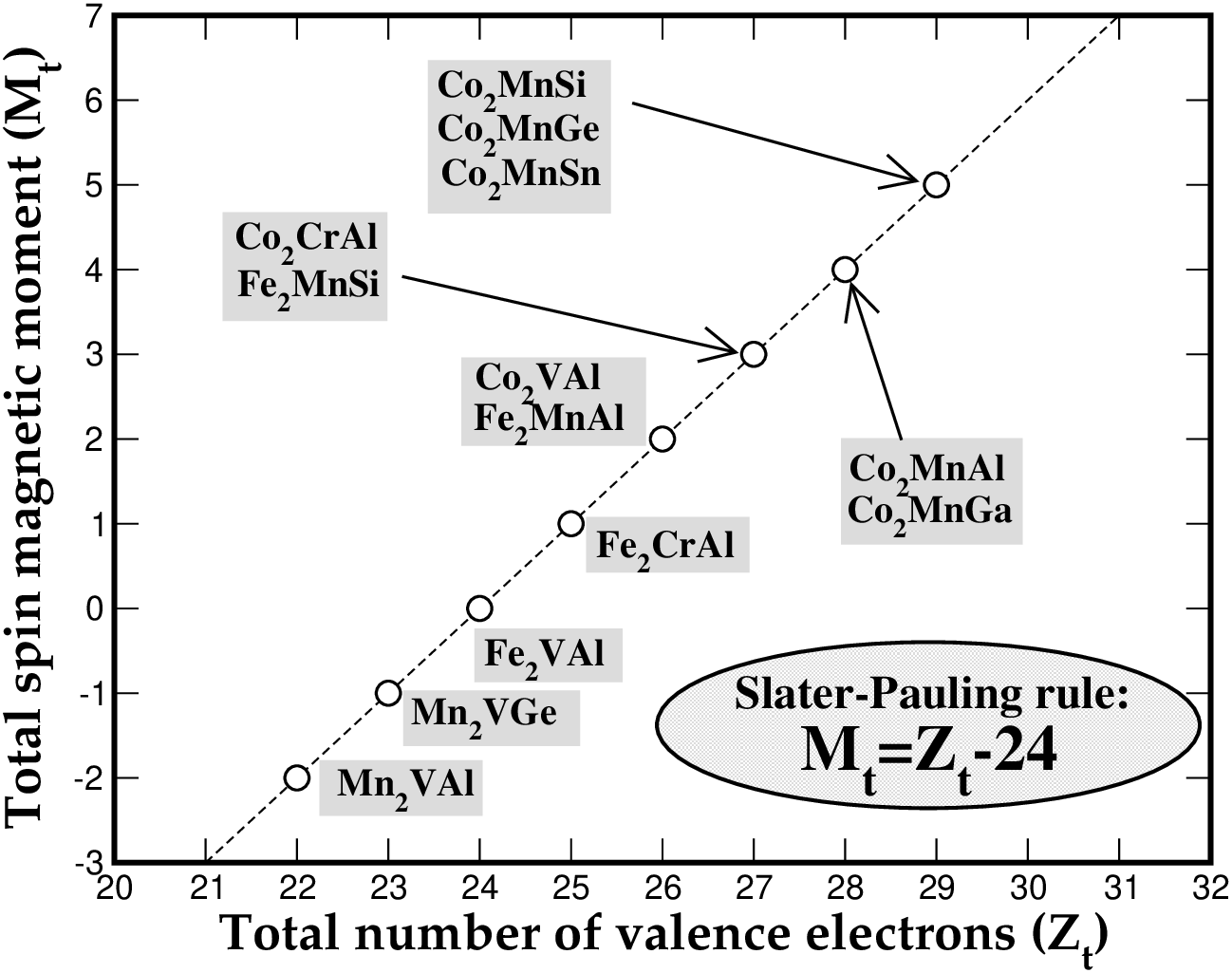}
\caption{Calculated total spin moment per unit cell in $\mu_B$ as
a function of  the total number $Z_t$ of valence electrons per
unit cell for the full Heusler alloys. The dashed line represents
the Slater-Pauling behavior.}
 \label{fig1}
\end{figure}

The electronic and magnetic properties of the full-Heusler
compounds are similar to the half-Heusler
compounds\cite{GalanakisHalf} with the additional complication of
the presence of 2 Co atoms per unit cell as in Co$_2$MnGe. In the
case of the semi-Heusler alloys there are exactly nine occupied
minority-spin electronic states; one of s character and three of
p-character provided by the sp atom and five bonding d-states
originating from the hybridization between the transition metal
atoms. In addition  to these bands, in the case of full-Heusler
alloys there are five states exclusively located at the Co sites
and near the Fermi level and the Fermi level is located among
these states so that three out of five are occupied and two of
them unoccupied leading to small energy gaps.\cite{GalanakisFull}
Now there are 12 minority-spin occupied states and the compounds
with 24 valence electrons like Fe$_2$VSb are semiconductors. If
vanadium is substituted by a higher-valent atom, spontaneous spin
polarization occurs, and the exchange splitting shifts the
majority states to lower energies. The extra electrons fill in
only majority states and the total spin moment per unit cell in
$\mu_\mathrm{B}$, $M_t$, which is the number of uncompensated
spins is the total number of valence electrons, $Z_t$, minus two
times the 12,: $M_t$=$Z_t$-24. This behavior is the so-called
Slater-Pauling behavior and e.g. half-metallic Co$_2$CrAl (27
valence electrons) has a total spin moment of 3 $\mu_\mathrm{B}$
and half-metallic Co$_2$MnSi (29 valence electrons) a spin moment
of 5 $\mu_\mathrm{B}$. This rule provides a direct connection
between the half-metallicity and the total spin moment which can
be easily determined experimentally. In figure \ref{fig1} we have
plotted the calculated total spin moments for several full-Heusler
compounds as a function of the total number of valence electrons.
The dashed line represents the Slater-Pauling rule of
half-metallic full Heusler alloys. Since 7 majority bands
(2$\times$Co $e_u$, 5$\times$Mn $d$) are unoccupied, the largest
possible moment is 7 $\mu_B$ and would occur if all majority
$d$-states were occupied. Of course it would be impossible to get
a compound with a total spin moment of 7 $\mu_B$ but even $M_t=6\
\mu_B$ is difficult to obtain. As it was shown by Wurmehl et
al.~\cite{Co2FeSi}, the on-site correlations in Co$_2$FeSi play a
critical role for this compound and calculations within the LDA+U
scheme, rather than the LDA, give a spin moment of 6 $\mu_B$. We
should also note here that  ferromagnetism is stabilized by the
inter-sublattice interactions between the Mn(Cr) and Co atoms and
between Co atoms belonging to different sublattices as shown by
first-principles calculations.\cite{GalanakisExchConst,Kurtulus}

Before closing this section we should discuss also the role of the
$sp$-elements in half-metallic Heusler alloys.  While the
$sp$-elements are not responsible for the appearance of the
minority gap, they are very important for the physical properties
of the Heusler alloys and their structural stability.  sp atoms
provide one s and three p bands per spin which lay very low in
energy and accommodate d-electrons of the transition metal atoms.
Thus the effective d-charge, which is accommodated by
transition-metal atomic d-hybrids, is  reduced stabilizing the
half-metallicity. These s- and p-states of the sp atom strongly
hybridize with the transition metal $d$-states and the charge in
these bands is delocalized and locally the sp atoms even lose
charge in favor of the transition metal atoms.

\section{Defects in full-Heuslers containing C\lowercase{o} and M\lowercase{n} \label{sec3} }

\begin{figure}
\includegraphics[width=\linewidth]{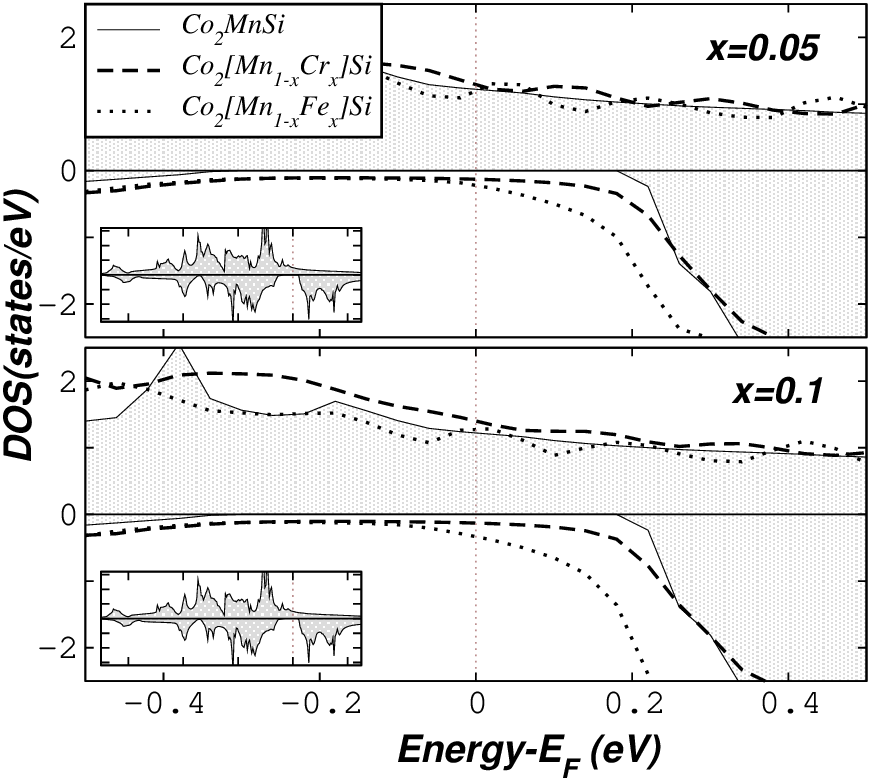}
\includegraphics[width=\linewidth]{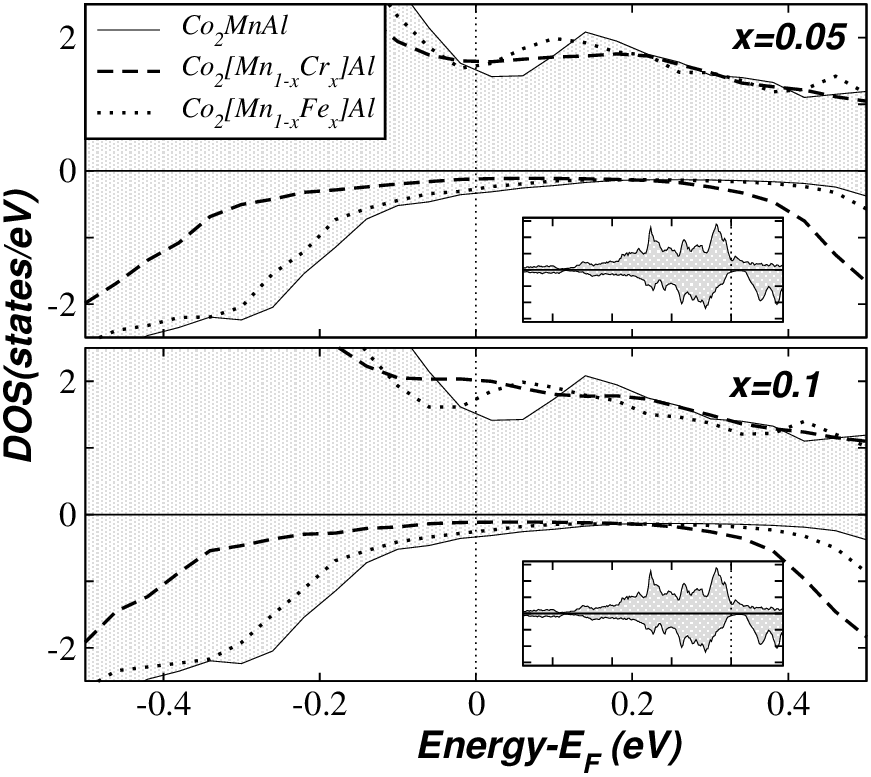}
\caption{Spin-resolved density of states (DOS) for the case of
Co$_2$[Mn$_{1-x}$Cr$_x$]Si and Co$_2$[Mn$_{1-x}$Fe$_x$]Si in the
upper panel, and  Co$_2$[Mn$_{1-x}$Cr$_x$]Al and
Co$_2$[Mn$_{1-x}$Fe$_x$]Al in the lower panel for two values of
the doping concentration $x$. DOS's are compared to the one of the
undoped Co$_2$MnSi and Co$_2$MnAl alloys. The zero of the energy
axis corresponds to the Fermi energy. Positive values of the DOS
correspond to the majority-spin (spin-up) electrons and negative
values to the minority-spin (spin-down) electrons. In the insets
we present the DOS for a wider energy range.\label{fig2}}
\end{figure}

In this section we discuss results on the defects in the case of
Co$_2$MnZ alloys where Z is Al and Si.\cite{APL,PRB-def} To
simulate the doping by electrons we substitute Fe for Mn while to
simulate the doping of the alloys with holes we substitute Cr for
Mn and we have considered the cases of moderate doping
substituting 5\%\ and 10\%\ of the Mn atoms and we present our
results in figure \ref{fig2} for the Co$_2$MnSi and Co$_2$MnAl
compounds. As discussed in the previous section the gap is created
between states located exclusively at the Co sites. The states low
in energy (around -6 eV below the Fermi level) originate from the
low-lying p-states of the sp atoms and the ones at around -9 eV
below the Fermi level are the s-states of the sp-atom.  The
majority-spin occupied states form a common Mn-Co band while the
occupied minority states are mainly located at the Co sites and
the minority unoccupied states at the Mn sites. The extra electron
in the the Co$_2$MnSi alloy occupies majority states leading to an
increase of the exchange splitting between the occupied majority
and the unoccupied minority states and thus to larger gap-width
for the Si-based compound. In the case of the Al-based alloy the
bonding and antibonding minority d-hybrids almost overlap and the
gap is substituted by a region of very small minority density of
states (DOS); we will call it a pseudogap. In both cases the Fermi
level falls within the gap (Co$_2$MnSi) or the pseudogap
(Co$_2$MnAl) and an almost perfect spin-polarization at the Fermi
level is preserved.

Doping the perfect ordered alloys with either Fe or Cr smoothens
the valleys and peaks along the energy axis. This is a clear sign
of the chemical disorder; Fe and Cr induce peaks at slightly
different places than the Mn atoms resulting to this smoothening
and as the doping increases this phenomenon becomes more intense.
The important detail is what happens around the Fermi level and in
what extent is the gap in the minority band affected by the
doping. So now we will concentrate only at the enlarged regions
around the Fermi level. The dashed lines represent the Cr-doping
while the dotted lines are the Fe-doped alloys. Cr-doping in
Co$_2$MnSi has only a marginal effect to the gap. Its width is
narrower with respect to the perfect compounds but overall the
compounds retain their half-metallicity. For Co$_2$MnAl the
situation is reversed with respect to the Co$_2$MnSi compound and
Cr-doping has significant effects on the pseudogap. Its width is
larger with respect to the perfect compound and becomes slightly
narrower as the degree of doping increases.

In the case of Fe-doping the situation is more complex. Adding
electrons to the system means that, in order to retain the perfect
half-metallicity, these electrons should occupy high-energy lying
antibonding majority states. This is energetically not very
favorable in the case of Co$_2$MnSi and for these moderate degrees
of doping a new shoulder appears in the unoccupied states which is
close to the right-edge of the gap; a sign of a large change in
the competition between the exchange splitting of the Mn majority
and minority states and of the Coulomb repulsion. In the case of
the 20\% Fe doping in Co$_2$MnSi (not shown here) this new peak
crosses the Fermi level and the Fermi level is no more exactly in
the gap but slightly above it. Further substitution should lead to
the complete destruction of the half-metallicity.\cite{GalaQuat}
Recent ab-initio calculations including the on-site Coulomb
repulsion (the so-called Hubbard $U$) have predicted that
Co$_2$FeSi is in reality half-metallic reaching a total spin
magnetic moment of 6 $\mu_B$ which is the largest known spin
moment for a half-metal.\cite{Co2FeSi,Wurmehl}
  Fe-doping on the other hand in Co$_2$MnAl almost does
not change the DOS around the Fermi level. The extra-electrons
occupy high-energy lying antibonding majority states but, since
Co$_2$MnAl has one valence electron less than Co$_2$MnSi,
half-metallicity remains energetically favorable and no important
changes occur upon Fe-doping and   further substitution of Fe for
Mn should retain the half-metallicity even for the Co$_2$FeAl
compound although LDA-based ab-initio calculations predict that
the limiting case of Co$_2$FeAl is almost
half-metallic.\cite{GalaQuat}

\section{Defects driven half-metallic ferrimagnetism \label{sec4}}

In the previous section we have examined the case of defects in
half-metallic ferromagnets.  Half-metallic ferrimagnetism (HMFi)
on the other hand is highly desirable since such compounds would
yield lower total spin moments than the corresponding
ferromagnets. Well-known HMFi are the perfect Heusler compounds
FeMnSb and Mn$_2$VAl.\cite{deGroot2} We will present in the first
part of this section another route to half-metallic ferrimagnetism
based on antisites created by the migration of Cr(Mn) atoms at Co
sites in the case of Co$_2$CrAl, Co$_2$CrSi, Co$_2$MnAl and
Co$_2$MnSi alloys. The ideal case for applications would be a
fully-compensated ferrimagnet,  also known as half-metallic
antiferromagnet (HMA),\cite{Leuken} since such a compound would
not give rise to stray flux and thus would lead to smaller energy
consumption in devices. In the secobd part of this section we
present a way to achieve HMA based on Co defects in half-metallic
ferrimagnets like Mn$_2$VAl.

\begin{figure}
\includegraphics[width=\linewidth]{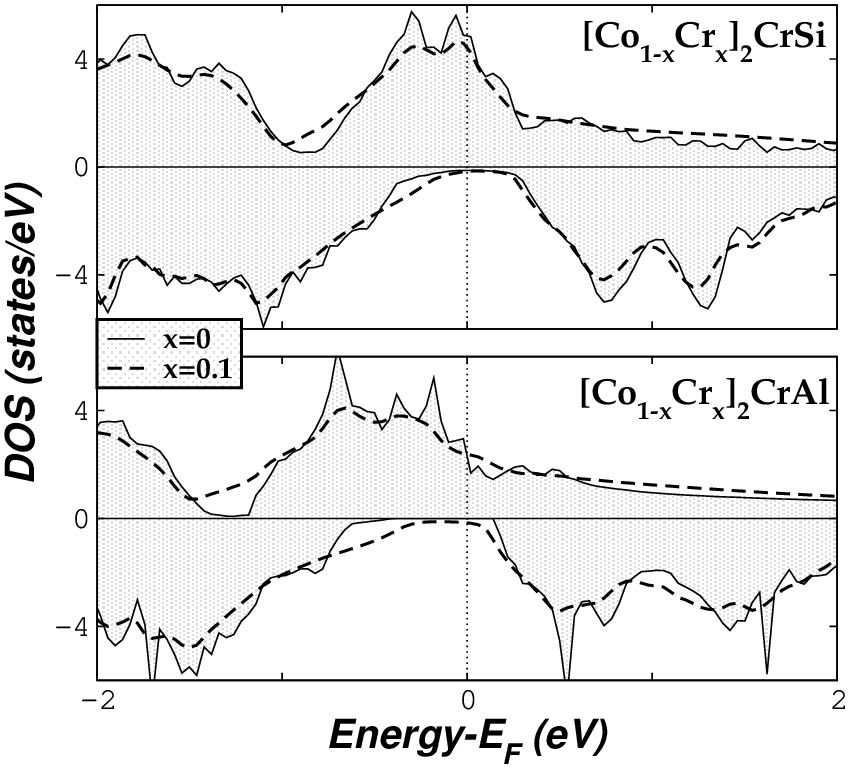}
\caption{Total density of states (DOS) as a function of the
concentration $x$  for the [Co$_{1-x}$Cr$_{x}$]$_2$CrAl (upper
panel) and [Co$_{1-x}$Cr$_{x}$]$_2$CrSi (lower panel) compounds.
\label{fig_HMFi}}
\end{figure}

\begin{table}
\caption{Atom-resolved spin magnetic moments for the
[Co$_{1-x}$Cr$_x$]$_2$CrAl, [Co$_{1-x}$Cr$_x$]$_2$CrSi,
[Co$_{1-x}$Mn$_x$]$_2$MnAl and [Co$_{1-x}$Mn$_x$]$_2$MnSi
compounds (moments have been scaled to one atom). The two last
columns are the total spin moment (Total) in the unit cell
calculated as $2\times [(1-x)*m^{Co}+x*m^{Cr \:\mathrm{or}\:
Mn(imp)}]+m^{Cr(Mn)}+m^{Al\:\mathrm{or}\:Si}$ and the ideal total
spin moment predicted by the Slater-Pauling rule for half-metals
(see section \ref{sec2}). With IMP  we denote the Cr(Mn) atoms
sitting at perfect Co sites. \label{table_HMFi} }
\begin{tabular}{lcccccc}\hline\noalign{\smallskip}
Compound & Co   & IMP & Cr/Mn & Al/Si & Total &Ideal \\
Co$_2$CrAl     &  0.73 & -- &  1.63 & -0.09&3.00  & 3.00 \\

[Co$_{0.95}$Cr$_{0.05}$]$_2$CrAl & 0.71 & -1.82 & 1.62 & -0.09 & 2.70 & 2.70 \\

[Co$_{0.9}$Cr$_{0.1}$]$_2$CrAl & 0.69 & -1.85 & 1.61 & -0.08 &
2.40 &  2.40 \\

Co$_2$CrSi   & 0.95 & -- & 2.17 & -0.06 & 4.00 &  4.00 \\

[Co$_{0.95}$Cr$_{0.05}$]$_2$CrSi & 0.93 & -1.26 & 2.12 & -0.06 & 3.70 & 3.70 \\

[Co$_{0.9}$Cr$_{0.1}$]$_2$CrSi & 0.91 & -1.26 & 2.07 & -0.05 & 3.40 & 3.40 \\

Co$_2$MnAl   &  0.68  & -- & 2.82 & -0.14 & 4.04 & 4.00 \\

[Co$_{0.95}$Mn$_{0.05}$]$_2$MnAl & 0.73 & -2.59 & 2.82 & -0.13 & 3.81 & 3.80 \\

[Co$_{0.9}$Mn$_{0.1}$]$_2$MnAl & 0.78 & -2.49 & 2.83 & -0.12 & 3.61 & 3.60 \\

Co$_2$MnSi &  0.98  & -- & 3.13 & -0.09 & 5.00 & 5.00\\

[Co$_{0.95}$Mn$_{0.05}$]$_2$MnSi & 0.99 & -0.95 & 3.09 & -0.08 & 4.80 & 4.80\\

[Co$_{0.9}$Mn$_{0.1}$]$_2$MnSi & 0.99 & -0.84 & 3.06 & -0.07 & 4.60& 4.60 \\
\noalign{\smallskip}\hline

\end{tabular}
\end{table}

We will start our discussion from  the Cr-based alloys and using
Co$_2$CrAl and Co$_2$CrSi as parent compounds we create a surplus
of Cr atoms which sit at the perfect Co sites. In figure
\ref{fig_HMFi}  we present the total density of states (DOS) for
the [Co$_{1-x}$Cr$_{x}$]$_2$CrAl  and [Co$_{1-x}$Cr$_{x}$]$_2$CrSi
alloys for concentrations $x$= 0 and 0.1, and in table
\ref{table_HMFi} we have gathered the spin moments for the two
compounds under study. We will start our discussion from the DOS.
The perfect compounds show a gap in the minority-spin band and the
Fermi level falls within this gap and thus the compounds are
half-metals. When the sp atom is Si instead of Al the gap is
larger due to the extra electron which occupies majority states of
the transition metal atoms\cite{GalanakisFull} and increases the
exchange splitting between the majority occupied and the minority
unoccupied states. This electron increases the Cr spin moment by
$\sim$0.5 $\mu_B$ and the moment of each Co atom by $\sim$0.25
$\mu_B$ about. The Cr and Co majority states form a common band
and the weight at the Fermi level is mainly of Cr character. The
minority occupied states are mainly of Co character. When we
substitute Cr for Co, the effect on the atomic DOS of the Co and
Cr atoms at the perfect sites is marginal. The DOS of the impurity
Cr atoms has a completely different form from the Cr atoms at the
perfect sites due to the different symmetry of the site where they
sit. But although Cr impurity atoms at the antisites induce
minority states within the gap, there is still a tiny gap and the
Fermi level falls within this gap keeping the half-metallic
character of the parent compounds.

The discussion above on the conservation of the half-metallicity
is confirmed when we compare the calculated total moments in table
\ref{table_HMFi}  with the values predicted by the Slater Pauling
rule for the ideal half-metals. Since Cr is lighter than Co,
substitution of Cr for Co decreases the total number of valence
electrons and the total spin moment should also decrease. This is
achieved due to the antiferromagnetic coupling between the Cr
impurity atoms and the Co and Cr ones at the ideal sites, which
would have an important negative contribution to the total moment
as confirmed by the results in table \ref{table_HMFi}. Thus the
Cr-doped alloys are half-metallic ferrimagnets and their total
spin moment is considerable smaller than the perfect half-metallic
ferromagnetic parent compounds; in the case of
[Co$_{0.8}$Cr$_{0.2}$]$_2$CrAl it decreases down to 1.8 $\mu_B$
from the 3 $\mu_B$ of the perfect Co$_2$CrAl alloy. Here we have
to mention that if also Co atoms migrate to Cr sites (case of
atomic swaps) the half-metallity is lost, as it was shown by Miura
et al. \cite{Miura}, due to the energy position of the Co states
which have migrated at Cr sites.

\begin{figure}
\begin{center}
\includegraphics[width=\linewidth]{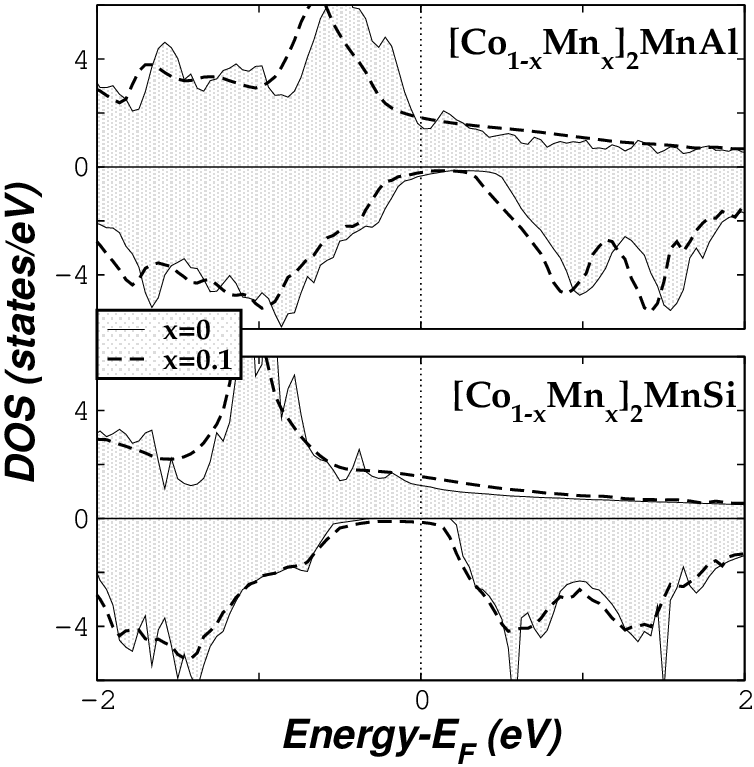}
\end{center}
\caption{Total density of states (DOS) around the gap region for
the [Co$_{1-x}$Mn$_{x}$]$_2$MnAl
 and  [Co$_{1-x}$Mn$_{x}$]$_2$MnSi alloys as a function of the
concentration $x$ : we denote $x=0$ with the solid line  and
$x=0.1$ with a dashed thick  line.} \label{fig_HMFi2}
\end{figure}

Similar phenomena to the discussion above occur in the
[Co$_{1-x}$Mn$_x$]$_2$MnZ compounds varying the sp atom, Z, which
is one of Al or Si.  We have taken into account five different
values for the concentration $x$; $x$= 0, 0.025, 0.05, 0.1, 0.2.
In figure \ref{fig_HMFi2} we have drawn the total density of
states (DOS) for both families of compounds under study and for
two different values of the concentration $x$: the perfect
compounds ($x$=0) and for one case with defects, $x$= 0.1. As
mentioned above and in agreement with previous  previous
electronic structure calculations on these compounds Co$_2$MnSi
compounds present a real gap and Co$_2$MnAl a
pseudogap.\cite{GalanakisFull,APL,Richter,JAP} When we create a
surplus of Mn atoms which migrate at sites occupied by Co atoms in
the perfect alloys, the gap persists and both compounds retain
their half-metallic character as occurs also for the Cr-based
alloys presented above. Especially for Co$_2$MnSi, the creation of
Mn antisites does not alter the width of the gap and the
half-metallicity is extremely robust in these alloys with respect
to the creation of Mn antisites. The atomic spin moments show
behavior similar to the Cr alloys as can be seen in table
\ref{table_HMFi} and the spin moments of the Mn impurity atoms are
antiferromagnetically coupled to the spin moments of the Co and Mn
atoms at the perfect sites resulting to the desired half-metallic
ferrimagnetism.

The ideal case of fully-compensated half-metallic ferrimagnet or
half-metallic antiferromagnetic (HMA) can be achieved by doping
with Co the Mn$_2$VAl and Mn$_2$VSi alloys which are well known to
be HMFi. The importance of this route stems from the existence of
Mn$_2$VAl in the Heusler $L2_1$ phase as shown by several
groups.\cite{itoh1,itoh2,itoh3} Each Mn atom has a spin moment of
around -1.5 $\mu_B$ and V atom a moment of about 0.9
$\mu_B$.\cite{itoh1,itoh2,itoh3}

\begin{figure}
\begin{center}
\includegraphics[width=\linewidth]{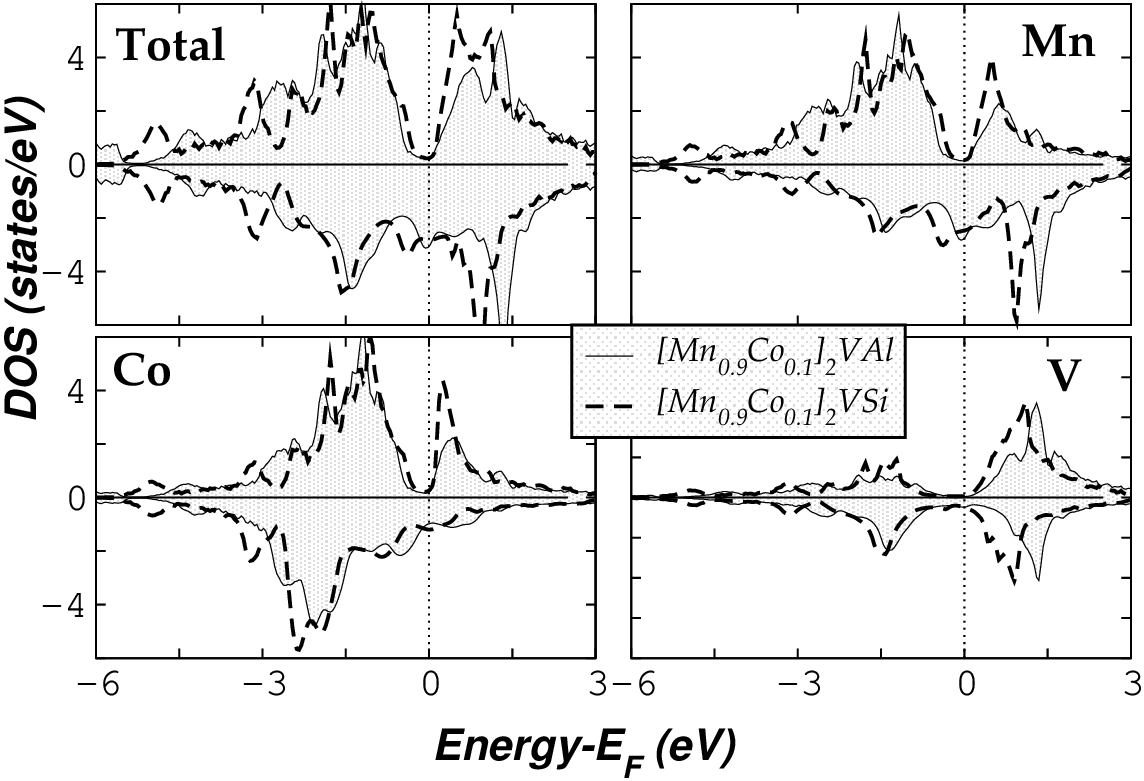}
\end{center}
\caption{Total and atom-resolved DOS for the
[Mn$_{0.9}$Co$_{0.1}$]$_2$VAl and [Mn$_{0.9}$Co$_{0.1}$]$_2$VSi
compounds. Note that the atomic DOS's have been scaled to one
atom. \label{fig_HMA} }
\end{figure}

 All theoretical studies on
Mn$_2$VAl agree on the half-metallic character with a gap at the
spin-up band instead of the spin-down band as for the other
half-metallic Heusler
alloys.\cite{GalanakisFull,Mn2VZ-2,Weht1,Weht2} Prior to the
presentation of our results we have to note that due to the
Slater-Pauling rule, these compounds with less than 24 valence
electrons have negative total spin moments and the gap is located
at the spin-up band. Moreover the spin-up electrons correspond to
the minority-spin electrons and the spin-down electrons to the
majority electrons contrary to the other Heusler
alloys.\cite{GalanakisFull} We have substituted Co for Mn in
Mn$_2$V(Al or Si) in a random way and in figure \ref{fig_HMA} we
present the total and atom-resolved density of states (DOS) in
[Mn$_{1-x}$Co$_x$]$_2$VAl (solid line) and
[Mn$_{1-x}$Co$_x$]$_2$VSi (dashed line) alloys for $x$=0.1. The
perfect compounds show a region of low spin-up DOS (we will call
it a ``pseudogap'') instead of a real gap. Upon doping the
pseudogap at the spin-up band persists and the quaternary alloys
keep the half-metallic character of the perfect Mn$_2$VAl and
Mn$_2$VSi compounds.  Co atoms are strongly polarized by the Mn
atoms since they occupy the same sublattice and they form Co-Mn
hybrids which afterwards interact with the V and Al or Si
states.\cite{GalanakisFull} The spin-up Co states form a common
band with the Mn ones and the spin-up DOS for both atoms has
similar shape. Mn atoms have less weight in the spin-down band
since they accommodate less charge than the heavier Co atoms.

In table \ref{table_HMA} we have gathered the total and
atom-resolved spin moments for all the Co-doped compounds as a
function of the concentration. We have gone up to a concentration
which corresponds to 24 valence electrons in the unit cell, thus
up to $x$=0.5 for the [Mn$_{1-x}$Co$_x$]$_2$VAl and x=0.25 for the
[Mn$_{1-x}$Co$_x$]$_2$VSi alloys. In the last column we have
included the total spin moment predicted by the Slater-Pauling
rule for the perfect half-metals. A comparison between the
calculated and ideal total spin moments reveals that all the
compounds under study are half-metals with very small deviations
due to the existence of a pseudogap instead of a real gap. Exactly
for 24 valence electrons the total spin moment vanishes as we will
discuss in the next paragraph. Co atoms have a spin moment
parallel to the V one and antiparallel to the Mn moment, and thus
the compounds retain their ferrimagnetic character. As we increase
the concentration of the Co atoms in the alloys, each Co  has more
Co atoms as neighbors, it hybridizes stronger with them and its
spin moment increases while the spin moment of the Mn atom
decreases (these changes are not too drastic). The sp atoms have a
spin moment antiparallel to the Mn atoms as already discussed in
reference \onlinecite{Mn2VZ}.

\begin{table}
\caption{Atom-resolved spin magnetic moments for the
[Mn$_{1-x}$Co$_x$]$_2$VAl and [Mn$_{1-x}$Co$_x$]$_2$VSi compounds
(moments have been scaled to one atom). The two last columns are
the total spin moment (Total) in the unit cell calculated as
$2\times [(1-x)*m^{Mn}+x*m^{Co}]+m^V+m^{Al\:or\:Si}$ and the ideal
total spin moment predicted by the Slater-Pauling rule for
half-metals. The lattice constants have been chosen 0.605 nm for
Mn$_2$VAl and 0.6175 for Mn$_2$VSi for which both systems are
half-metals (see reference \onlinecite{Mn2VZ}) and have been kept
constant upon Co doping. } \label{table_HMA}
 \begin{tabular}{lcccccc}
 \hline\noalign{\smallskip}

 Compound & Mn   & Co & V & Al & Total &Ideal \\
Mn$_2$VAl     & -1.57  & -- & 1.08  & 0.06 & -2.00 & -2.0 \\

[Mn$_{0.9}$Co$_{0.1}$]$_2$VAl  & -1.56 & 0.340 & 1.07 & 0.07 & -1.60 & -1.6 \\

[Mn$_{0.7}$Co$_{0.3}$]$_2$VAl  & -1.48 & 0.46 & 0.95 & 0.05 & -0.80 & -0.8 \\

[Mn$_{0.5}$Co$_{0.5}$]$_2$VAl & -1.39 & 0.59 & 0.78 & 0.02 &
$\sim$0 & 0\\ \hline

Compound & Mn   & Co & V & Si & Total &Ideal \\

Mn$_2$VSi  &  -0.96 & -- & 0.86 & 0.06&-1.00 &  -1.0 \\

[Mn$_{0.9}$Co$_{0.1}$]$_2$VSi & -0.93 & 0.82 & 0.85 & 0.05 & -0.60 & -0.6 \\

[Mn$_{0.75}$Co$_{0.25}$]$_2$VSi &-0.90 & 0.94 & 0.84 & 0.04 & $\sim$0 & 0 \\
\noalign{\smallskip}\hline
\end{tabular}

\end{table}

The most interesting point in this substitution procedure is
revealed when  we increase the Co concentration to a value
corresponding to 24 valence electrons in the unit cell, thus  the
[Mn$_{0.5}$Co$_{0.5}$]$_2$VAl and [Mn$_{0.75}$Co$_{0.25}$]$_2$VSi
alloys. The Slater-Pauling rule predicts for these compounds a
zero total spin moment in the unit cell and  the electrons
population is equally divided between the two spin-bands. Our
first-principles calculations reveal that this is actually the
case. The interest arises from the fact that although the total
moment is zero, these two compounds are made up from  strongly
magnetic components. Mn atoms have a mean spin moment of
$\sim$-1.4 $\mu_B$ in [Mn$_{0.5}$Co$_{0.5}$]$_2$VAl and $\sim$-0.9
$\mu_B$ in [Mn$_{0.75}$Co$_{0.25}$]$_2$VSi. Co and V have spin
moments antiferromagnetically coupled to the Mn ones which for
[Mn$_{0.5}$Co$_{0.5}$]$_2$VAl are  $\sim$0.6 and $\sim$0.8
$\mu_B$, respectively, and for [Mn$_{0.75}$Co$_{0.25}$]$_2$VSi
$\sim$0.9 and $\sim$0.8 $\mu_B$. Thus these two compounds are
half-metallic fully-compensated ferrimagnets or as they are best
known in literature half-metallic antiferromagnets.

\section{Summary and conclusions \label{sec5}}

In this chapter we have reviewed our results on the defects in
half-metallic full-Heusler alloys. Firstly we have presented a
short overview of the electronic and magnetic properties of the
half-metallic full-Heusler alloys and have discussed in detail the
Slater-Pauling behavior of these alloys (the total spin moment
scales linearly with the total number of valence electrons as a
result of the half-metallicity). We have also studied the effect
of doping on the magnetic properties of the Co$_2$MnAl(Si)
full-Heusler alloys. Doping simulated by the substitution of Cr
and Fe for Mn overall keeps the half-metallicity and has   little
effect on the half-metallic properties of Co$_2$MnSi   and
Co$_2$MnAl compounds.

Afterwards, we have studied the effect of defects-driven
appearance of half-metallic ferrimagnetism in the case of the
Co$_2$Cr(Mn)Al(Si) Heusler alloys.  More precisely, based on
first-principles calculations we have shown that when we create
Cr(Mn) antisites at the Co sites, these impurity Cr(Mn) atoms
couple antiferromagnetically with the Co and the Cr(Mn) atoms at
the perfect sites while keeping the half-metallic character of the
parent compounds. The ideal case of half-metallic
fully-compensated ferrimagnets (also known as half-metallic
antiferromagnets) can be achieved by doping of the half-metallic
ferrimagnets Mn$_2$VAl and Mn$_2$VSi. Co substitution for Mn keeps
the half-metallic character of the parent compounds and when the
total number of valence electrons reaches the 24, the total spin
moment vanishes as predicted by the Slater-Pauling rule. Defects
are a promising alternative way to create robust half-metallic
ferrimagnets, which are crucial for magnetoelectronic
applications.

\end{document}